\documentclass[preprintnumbers, prd, showpacs, floatfix,twocolumn,
preprintnumbers, letterpaper, superscriptaddress,nofootinbib]{revtex4}
\usepackage{amsfonts}
\usepackage{amsmath}
\usepackage{amssymb,epsf}
\usepackage{latexsym}
\usepackage{graphicx,epsfig}
\usepackage{amssymb}
\usepackage{subfigure}
\usepackage{epstopdf}
\usepackage[colorlinks=true,citecolor=blue,linkcolor=blue,urlcolor=black]{hyperref}

\begin{document}

\title{Higher-dimensional thin-shell wormholes in third-order Lovelock
gravity}
\author{Mohammad Reza Mehdizadeh}
\email{mehdizadeh.mr@uk.ac.ir}
\affiliation{Department of Physics, Shahid Bahonar University, P.O. Box 76175, Kerman,
Iran}
\affiliation{Research Institute for Astronomy and Astrophysics of Maragha (RIAAM), P.O.
Box 55134-441, Maragha, Iran}
\author{Mahdi Kord Zangeneh}
\email{mkzangeneh@shirazu.ac.ir}
\affiliation{Physics Department and Biruni Observatory, College of Sciences, Shiraz
University, Shiraz 71454, Iran}
\author{Francisco S. N. Lobo}
\email{flobo@cii.fc.ul.pt}
\affiliation{Instituto de Astrof\'isica e Ci\^encias do Espa\c{c}o, Universidade de
Lisboa, Faculdade de Ci\^encias, Campo Grande, PT1749-016 Lisboa, Portugal}
\date{\today }

\begin{abstract}
In this work, we explore asymptotically flat charged thin-shell wormholes of
third order Lovelock gravity in higher dimensions, taking into account the
cut-and-paste technique. Using the generalized junction conditions, we
determine the energy-momentum tensor of these solutions on the shell, and
explore the issue of the energy conditions and the amount of normal matter
that supports these thin-shell wormholes. Our analysis shows that for
negative second order and positive third-order Lovelock coefficients, there
are thin-shell wormhole solutions that respect the weak energy condition. In
this case, the amount of normal matter increases as the third-order Lovelock
coefficient decreases. We also find novel solutions which possess specific
regions where the energy conditions are satisfied for the case of a positive
second order and negative third-order Lovelock coefficients. Finally, a
linear stability analysis in higher dimensions around the static solutions
is carried out. Considering a specific cold equation of state, we find a
wide range of stability regions.
\end{abstract}

\maketitle

\section{INTRODUCTION}

Wormholes are topologically nontrivial objects which connect two separate
and distinct spacetime regions \cite{MT,MTY}. A fundamental ingredient in
wormhole physics is the flaring-out condition of the throat, which in
General Relativity (GR) entails the violation of the null energy condition
(NEC). Matter that violates the NEC is denoted exotic matter. In fact,
wormholes violate all of the pointwise and averaged energy condition \cite%
{MTY}. Due to the problematic nature of these violations, it is useful to
minimize the exoticity problem of the matter threading and sustaining the
wormhole. For instance, traversable wormholes supported by arbitrarily small
quantities of exotic matter \cite{viskardad} or by matter respecting the
energy conditions \cite{MartRich,Deh1,wecsat,Harko:2013yb} have been
investigated.

An interesting kind of traversable wormhole is the thin-shell variety
constructed by surgically grafting together two spacetimes to form a
geodesically complete manifold with a shell placed at the junction
interface. In these types of wormholes, matter is concentrated on the
wormhole throat, which coincides with the junction boundary, and thus
minimizes the usage of exotic matter. These solutions were initially
presented in \cite{Vis,VisP} and have been extensively investigated in the
literature \cite{thin,lemo11}. In GR, it was shown that thin-shell wormholes
inevitably violate the weak energy condition (WEC) and, consequently,
various attempts have been undertaken, to alleviate this problem, in
alternative theories of gravity, for instance, in dilatonic gravity \cite%
{bej}, Brans-Dicke theory \cite{yueca} and DGP gravity \cite{cyl}. An
advantage of the thin-shell construction is that one may consider a
linearized stability analysis around the static solution, while preserving
the symmetric configuration, and analyse the dynamical evolution of the
thin-shell \cite{brad}. The generalization of the stability analysis to the
case of higher-dimensional wormholes has also been extensively investigated 
\cite{rah1}. For instance, a linearized stability analysis in the presence
of a dilaton, axion, phantom and other types of matter and in cylindrical
symmetry have also been studied \cite{eir}.

Since string theory proposes higher dimensional gravitational objects, it is
of interest to explore the possibility of higher dimensional thin-shell
wormhole solutions. One of the most interesting higher dimensional
alternative theories of gravity which leads to second order field equations
for the metric is described by the Lovelock action \cite{lovee}. In this
context, thin shell wormholes of second order Lovelock theory called
Gauss-Bonnet theory in $5$-dimensions and higher dimensions were extensively
analysed \cite{mazv}. In the context of third order Lovelock gravity,
asymptotically flat and charged thin-shell wormholes of Lovelock gravity in
seven dimensions were constructed, using the cut-and-paste technique, and
the generalized junction conditions were applied in order to calculate the
energy-momentum tensor of these wormholes on the shell \cite{meh}. It was
found that for negative second order and positive third order Lovelock
coefficients, there are thin-shell wormholes that respect the WEC in
specific regions. In this case, the amount of normal matter decreases as the
third order Lovelock coefficient increases. For positive second and third
order Lovelock coefficients, the WEC is violated and the amount of exotic
matter decreases as the charge increases. Furthermore, a linear stability
analysis was analysed against a symmetry preserving perturbation, and the
stability of these wormhole were studied.

In this work, we extend the analysis outlined in \cite{meh}, by providing
the explicit form of the surface energy-momentum tensor and generalize the
analysis to arbitrary dimensions. More specifically, we present the explicit
form of the energy-momentum tensor in terms of the extrinsic curvature and
Riemann tensor for all higher dimensions, and find specific solutions,
including novel solutions which possess specific regions where the energy
conditions are satisfied for the particular cases of positive second order
and negative third-order Lovelock coefficients. Furthermore, we present the
generalized junction conditions, and analyse the stability analysis for the
solutions found, for arbitrary dimensions. Thus, we study the
effects of the third order Lovelock term in higher dimensionality, its role in the
satisfaction of the WEC and the stability of thin shell wormhole solutions in
higher dimensions.

This paper is organised in the following manner: In Section \ref{Sol}, we
present the action of third order Lovelock gravity in the presence of an
electromagnetic field, and construct a thin-shell wormhole in Lovelock gravity, by presenting explicitly the surface energy-momentum tensor components on the shell.
In particular, to complement the above Section \ref{Sol}, the specific 
$(n+1)$-dimensional static solutions, that are to be used throughout the work, are outlined in Appendix \ref{app:I}. In Appendix \ref{Thin}, we also introduce the junction conditions in Lovelock gravity, in order to construct the thin-shell wormhole. In Section \ref{Exot}, we analyse the
profile of the surface stresses and consider the issue of the energy
conditions and the amount of normal matter supporting the thin-shell
wormhole. In Section \ref{Stab}, we study the stability of the thin-shell
wormholes under small perturbations preserving the symmetry of the wormhole
configuration, and explore the dynamical evolution of the shell. Finally, in
Section \ref{Concl}, we conclude.

\section{ACTION AND THIN-SHELL CONSTRUCTION}
\label{Sol}


The action of third order Lovelock gravity in the presence of an
electromagnetic field may be written as%
\begin{equation}
I=\int d^{n+1}x\sqrt{-g}\left( \mathcal{L}_{1}+\alpha _{2}\mathcal{L}%
_{2}+\alpha _{3}\mathcal{L}_{3}-F_{\mu \nu }F^{\mu \nu }\right) \,,
\label{Act1}
\end{equation}%
where $\alpha _{2}$ and $\alpha _{3}$ are the second (Gauss-Bonnet) and
third order Lovelock coefficients. The term $\mathcal{L}_{1}=R$ is the
Einstein-Hilbert Lagrangian, $\mathcal{L}_{2}$ is the Gauss-Bonnet
Lagrangian, and $\mathcal{L}_{3}$ is the third order Lovelock Lagrangian, which are defined in Appendix \ref{app:I}.

The electromagnetic tensor is defined as $F_{\mu \nu
}=\partial _{\mu }A_{\nu }-\partial _{\nu }A_{\mu }$, where the vector
potential $A_{\mu }$ is given by 
\begin{equation}
A_{\mu }=\sqrt{\frac{(n-1)}{2(n-2)}}\frac{q}{r^{n-2}}\delta _{\mu }^{t}.
\label{A}
\end{equation}%

The $(n+1)$-dimensional static solution of action (\ref{Act1}) is given in
the form%
\begin{equation}
ds^{2}=-f(r)dt^{2}+\frac{dr^{2}}{f(r)}+r^{2}d\Sigma _{n-1}^{2},  \label{met}
\end{equation}%
where $d\Sigma _{n-1}^{2}$ is the metric of a $(n-1)$-dimensional unit
sphere and the metric function $f(r)$ is given by%
\begin{equation}
f(r)=1-r^{2}\psi (r),  \label{f(r)}
\end{equation}
where the solutions for the function $\psi(r)$ are outlined in the Appendix \ref{app:I}.


To construct a thin-shell wormhole in third order Lovelock gravity, we use
the well-known cut-and-paste technique. The junction conditions in Lovelock gravity are given in Appendix \ref{Thin}, where we refer the reader for more details.
Taking two copies of the
asymptotically flat solutions of Lovelock gravity given by Eqs. (\ref{met})
and (\ref{f(r)}) and solutions of Eq. (\ref{Cubic}) and removing from each
manifold the $(n+1)$-dimensional region described by%
\begin{equation}
\Omega _{\pm }=\left\{ r_{\pm }\leq a,a>r_{h}\right\},
\end{equation}%
we are left with two geodesically incomplete manifolds with the following
timelike hypersurface as boundaries 
\begin{equation}
\Sigma _{\pm }=\left\{ r_{\pm }=a,a>r_{h}\right\} .
\end{equation}%
Now identifying these two boundaries, $\Sigma _{+}=\Sigma _{-}=\Sigma $, we
obtain a geodesically complete manifold containing the two asymptotically
flat regions $\Omega _{+}$ and $\Omega _{-}$ which are connected by a
wormhole. The throat of the wormhole is located at $\Sigma $ with the
following metric%
\begin{equation}
ds_{\Sigma }^{2}=-d\tau ^{2}+a^{2}(\tau )d\Omega _{n-1}^{2},
\end{equation}%
where $\tau $\ is the proper time along the hypersurface $\Sigma $ and $%
a(\tau )$ is the radius of the throat. In order to study the stability of
the thin-shell wormhole constructions, we allow the radius of the throat to
be a function of the proper time. In addition to this, note that all the
matter is concentrated on the wormhole throat, $\Sigma $.

To analyze such a thin-shell configuration, we need to use the modified
junction conditions. Considering the induced coordinates on $\Sigma $, $\xi
^{a}=(\tau ,\theta ^{i};$ $i=1...n)$, the extrinsic curvatures associated
with the two sides of the shell are given by 
\begin{equation}
{\mathcal{K}}_{ab}^{\pm }=-n_{\rho }^{\pm }\left( \frac{\partial ^{2}X^{\rho
}}{\partial \xi ^{a}\partial \xi ^{b}}+\Gamma _{\mu \nu }^{\rho }\frac{%
\partial X^{\mu }}{\partial \xi ^{a}}\frac{\partial X^{\nu }}{\partial \xi
^{b}}\right) _{r=a},
\end{equation}%
where the normal vector $n_{\rho }^{\pm }$ ($n_{\rho }n^{\rho }=1$) to the
surface $\Sigma $ in $\mathcal{M}$ is defined by 
\begin{equation}
n_{\gamma }^{\pm }=\pm \left\vert g^{\mu \nu }\frac{\partial \mathcal{G}}{%
\partial X^{\mu }}\frac{\partial \mathcal{G}}{\partial X^{\nu }}\right\vert 
\frac{\partial \mathcal{G}}{\partial X^{\gamma }} \,,
\end{equation}%
where $\mathcal{G}(r,\tau )$ is the equation of the boundary $\Sigma $,
given by 
\begin{equation}
\mathcal{G}(r,\tau )=r-a(\tau )=0.
\end{equation}

After some algebraic manipulations, in an orthonormal basis $\{e_{\hat{\tau}%
},e_{\hat{\imath}};$ $i=1...n-1\}$, the components of the extrinsic
curvature tensor are given by 
\begin{eqnarray}
\mathcal{K}_{~\hat{\tau}}^{\hat{\tau}} &=&\frac{\Gamma }{\Delta },
\label{Kia} \\
\mathcal{K}_{\hat{\jmath}}^{\hat{\imath}} &=&\frac{\Delta }{a}~\delta _{\hat{%
\jmath}}^{\hat{\imath}}  \label{Ki}
\end{eqnarray}%
with 
\begin{equation*}
\Gamma =\ddot{a}+\frac{f^{\prime }\left( a\right) }{2}, \qquad \Delta = 
\sqrt{\dot{a}^{2}+f\left( a\right) },
\end{equation*}%
where the prime and the overdot denote derivatives with respect to $a$ and $%
\tau $, respectively. Equations (\ref{Kia}) and (\ref{Ki}) show that the
form of the energy-momentum tensor on the shell is ${S}_{~\hat{b}}^{\hat{a}%
}=~\mbox{diag}~(-\sigma ,p~\delta _{\hat{\jmath}}^{\hat{\imath}})$, where $%
\sigma $ is the surface energy density and $p$ is the transverse pressure.

Now using the junction condition (\ref{explicit junction ij}), the
components of the energy-momentum tensor on the shell may be written as 
\begin{eqnarray}
\sigma &=&-S_{\tau }^{\tau }=-\frac{\Delta }{4\pi }\bigg\{\frac{(n-1)}{a}-%
\frac{2\tilde{\alpha}_{2}(n-1)}{3a^{3}}  \notag \\
&&\times \left[\Delta ^{2}-3(1+\dot{a}^{2})\right]+\frac{(n-1)\tilde{\alpha}%
_{3}}{5a^{5}}  \notag \\
&&\times \left[15(1+\dot{a}^{2})^{2}-10\Delta ^{2}(1+\dot{a}^{2})+3\Delta
^{4}\right]\bigg\},  \label{gesig0}
\end{eqnarray}%
\begin{eqnarray}
p &=&S_{i}^{i}=\frac{1}{8\pi }\bigg\{\frac{2\Gamma }{\Delta }+\frac{%
2(n-2)\Delta }{a}  \notag \\
&&-\frac{4\tilde{\alpha}_{2}}{3a^{2}}\Big\{3\Gamma \Delta -\frac{3\Gamma }{%
\Delta }\left( 1+\dot{a}^{2}\right) +\frac{\Delta ^{3}(n-4)}{a}  \notag \\
&&-\frac{6\Delta }{a} \left[ \frac{(n-4)}{2}\left( 1+\dot{a}^{2}\right) +%
\ddot{a}a \right]\Bigg\}+\frac{2\tilde{\alpha}_{3}}{5a^{4}}  \notag \\
&&\times \Bigg\{\frac{15\Gamma }{\Delta }\left( 1+\dot{a}^{2}-\Delta
^{2}\right) ^{2}+20\ddot{a}\Delta \left[ 3\left( 1+\dot{a}\right) -\Delta
^{2}\right]  \notag \\
&&+\frac{(n-6)\Delta }{5a}[15(1+{\dot{a}}^{2}-\Delta ^{2})^{2}  \notag \\
&&+20\left( 1+\dot{a}^{2}\right) \left( 3(1+\dot{a}^{2})-\Delta ^{2}\right) ]%
\big\}\Big\} \,,  \label{pre}
\end{eqnarray}
respectively. From the above equations we see that $\sigma $ and $p$ are
expressed in terms of the throat radius $a(\tau )$, the first and second
derivatives of $a(\tau )$ and the metric function $f(a)$. Note that the
surface energy density and transverse pressure satisfy the energy
conservation equation, given by 
\begin{equation}
\frac{d}{d\tau }\left( \sigma a^{(n-1)}\right) +p\frac{d}{d\tau }\left(
a^{(n-1)}\right) =0.  \label{eqco}
\end{equation}
The first term in Eq. (\ref{eqco}) represents the internal energy change of
the shell and the second term shows the work by internal forces of the shell.


\section{PROFILE OF THE SURFACE ENERGY-MOMENTUM ON THE THIN-SHELL}
\label{Exot}

In this section, we consider the issue of the energy conditions and the
amount of normal matter that supports the thin-shell wormhole. The WEC is
defined as $T_{\mu \nu }V^{\mu }V^{\nu }\geq 0$ for every nonspacelike
vector $V_{\mu }$. The WEC on the thin shell is satisfied provided that $%
\sigma \geq 0 $, $\sigma +p\geq 0$. We restrict our analysis to static
configurations with $a=a_{0}$ and $\dot{a}=\ddot{a}=0$. For the static
configuration, Eqs. (\ref{gesig0}) and (\ref{pre}) reduce to 
\begin{eqnarray}
\sigma _{0} &=&-\frac{(n-1)}{8\pi a_{0}\sqrt{f_{0}}}\left\{ 2f_{0}+\frac{4%
\tilde{\alpha}_{2}}{3a_{0}^{2}}f_{0}(3-f_{0})\right.  \notag \\
&&\left. +\frac{2\tilde{\alpha}_{3}}{5a_{0}^{4}}%
f_{0}(15-10f_{0}+3f_{0}{}^{2})\right\} ,  \label{s11}
\end{eqnarray}%
\begin{eqnarray}
\sigma _{0}+p_{0}&=&\frac{1}{8\pi a_{0}\sqrt{f_{0}}}\Big\{%
(-2f_{0}+a_{0}f_{0}^{\prime })  \notag \\
&& +\frac{2\tilde{\alpha}_{2}}{3a_{0}^{2}}\left[3a_{0}f_{0}^{\prime
}(1-f_{0}) -6f_{0}(3-f_{0})\right]  \notag \\
&& +\frac{\tilde{\alpha}_{3}}{a_{0}^{4}}[3a_{0}f_{0}^{\prime
}(1-2f_{0}+f_{0}^{2})  \notag \\
&& -2f_{0}(15-10f_{0}+3f_{0}{}^{2})]\Big\},  \label{po}
\end{eqnarray}%
respectively, where $f_{0}=f(a_{0})$ and $f_{0}^{\prime }=f^{\prime }(a_{0})$%
. In contrast to the case of general relativistic thin-shell wormholes for
which $\sigma_{0}<0$, thus violating the WEC \cite{rah1}, in
higher-dimensional thin-shell wormholes one can have, in principle, normal
matter on the shell. In order to investigate the profile of the surface
energy-momentum on the thin-shell, we find the amount of matter on the
shell, which is given by 
\begin{equation}
\digamma =\int drd\Omega _{n-1}[\sigma _{0}\delta (r-a_{0})+p_{r}].
\end{equation}%
For our case, the shell does not exert radial pressure, i.e., $p_{r}=0$, and
therefore the amount of matter on the shell is%
\begin{eqnarray}
\digamma &=&\frac{2\pi ^{\frac{n}{2}}a_{0}^{n-1}\sigma _{0}}{\Gamma (\frac{n%
}{2})}  \notag \\
&=&\frac{4\pi ^{\frac{n}{2}}(n-1)}{\Gamma (\frac{n}{2})}\sqrt{f_{0}}\; \bigg[%
-a_{0}^{(n-2)}+\frac{2\tilde{\alpha}_{2}a_{0}^{(n-4)}}{3}(f_{0}-3)  \notag \\
&&+\frac{\tilde{\alpha}_{3}a_{0}^{(n-6)}}{5}\left(-15+10f_{0}-3f_{0}{}^{2}%
\right)\bigg],  \label{FF}
\end{eqnarray}%
where $f_{0}$ is the value of $f(r)$ at $r=a_{0}$. In Einstein gravity, $%
\tilde{\alpha}_{2}=\tilde{\alpha}_{3}=0$, the matter is exotic both for
Reissner-Nordstrom $(Q\neq 0)$ and Schwarzschild $(Q=0)$ thin-shell
wormholes as one can see from Eq. (\ref{FF}) as $\sigma _{0}<0$ (Eq. (\ref%
{s11})). In Gauss-Bonnet gravity, $\alpha _{3}=0$ with $\alpha _{2}\leq 0 $,
one may have thin-shell wormholes supported by normal matter \cite{mazv}.

Now, we investigate the condition that thin-shell wormholes may be supported
by normal matter in third order Lovelock gravity. It is clear from Eq. (\ref%
{FF}) that the sign of $\digamma $ depends on the sign of $\sigma _{0}$.

\subsection{Specific case: $\tilde{\alpha}_{2}^{2}=3\tilde{\alpha}_{3}=\alpha ^{2}$}

For the special case $\tilde{\alpha}_{2}^{2}=3\tilde{\alpha}_{3}=\alpha ^{2}$%
, Eq. (\ref{s11}) takes the form 
\begin{eqnarray}
\sigma _{0} &=&-\frac{(n-1)}{60\pi a_{0}^{5}}\big[3\alpha
^{2}f_{0}^{2}-10\alpha (\alpha +a_{0}^{2})f_{0}  \notag \\
&&+15(\alpha ^{2}+2\alpha a_{0}^{2}+a_{0}^{4})\big].  \label{ssigm0}
\end{eqnarray}%
The sign of the square brackets in the above equation determines the sign of 
$\sigma _{0}$. This is a quadratic equation for the metric function $f_{0}$
and one can easily obtain the discriminant of the quadratic equation which
is equal to $-80\alpha ^{2}(\alpha +a_{0}^{2})^{2}$. Since the discriminant
is negative for all values of $\alpha $, the sign of the brackets is always
positive and therefore $\sigma _{0}<0$. In other words, the matter
supporting the thin-shell wormhole is exotic for any dimensions.

\subsection{Specific case: $\alpha_{2}>0$ and $\alpha_{3}>0$}

For the general solutions of third order Lovelock gravity we first consider
the case where $\tilde{\alpha}_{2}$ and $\tilde{\alpha}_{3}$ are positive.
In this case, since the metric function is asymptotically flat ($0<f_{0}<1$)
for $a\geqslant r_{h}$, and therefore the factors $3-f_{0}$ and $%
15-10f_{0}+3f_{0}^{2}$ are positive, the bracket in Eq. (\ref{s11}) is
always positive. Therefore, the matter on the throat is exotic in the case
of $\tilde{\alpha}_{2},\tilde{\alpha}_{3}>0$ as in this case we have $\sigma
_{0}<0$ (or $\digamma <0$ ).

\subsection{Specific case: $\alpha_{2}<0$ and $\alpha_{3}>0$}

\begin{figure*}[h]
\centering {%
\subfigure[~$\alpha_{3}=2$]{
   \label{k0}\includegraphics[width=.47\textwidth]{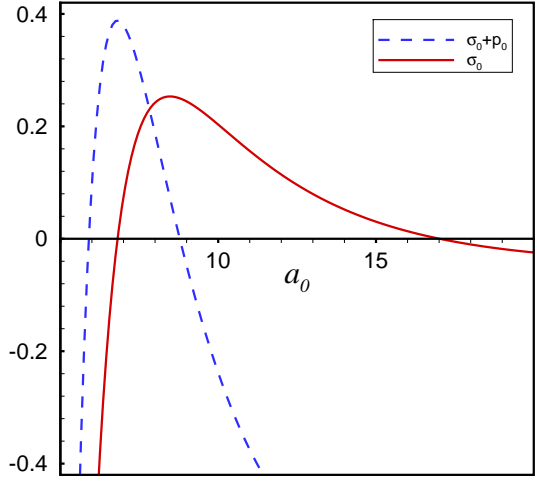}\qquad}} 
\subfigure[~$\alpha_{3}=1$]{
   \label{k1}\includegraphics[width=.47\textwidth]{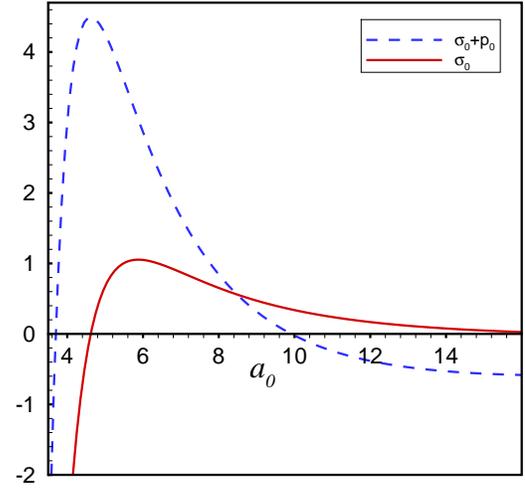}}
\caption{Specific case: $\protect\alpha _{2}<0$ and $\protect\alpha _{3}>0$. 
$\protect\sigma _{0}$ (continuous) and $\protect\sigma _{0}+p_{0}$ (dashed)
versus $a_{0}$ for $\protect\alpha _{2}=-2$, $m=100$, $q=3000$ and $n=9$.
These figures depict that for $\protect\alpha _{2}<0$ and $\protect\alpha %
_{3}>0$, the WEC can be satisfied. The plots also show that normal matter
for $\protect\alpha _{2}<0$ decreases as $\protect\alpha _{3}>0$ increases.}
\label{f34}
\end{figure*}

\begin{figure*}[h]
\centering
{%
\subfigure[$\alpha_{2}=-2$]{\label{k3}
\includegraphics[width=.47\textwidth]{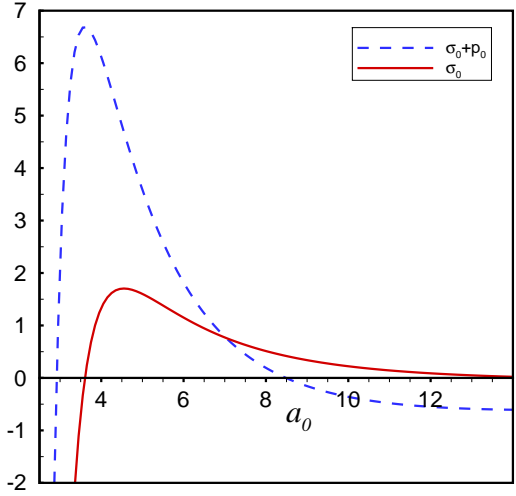}\qquad}} \subfigure[~$%
\alpha_{2}=-1.5$]{\label{k4}\includegraphics[width=.47\textwidth]{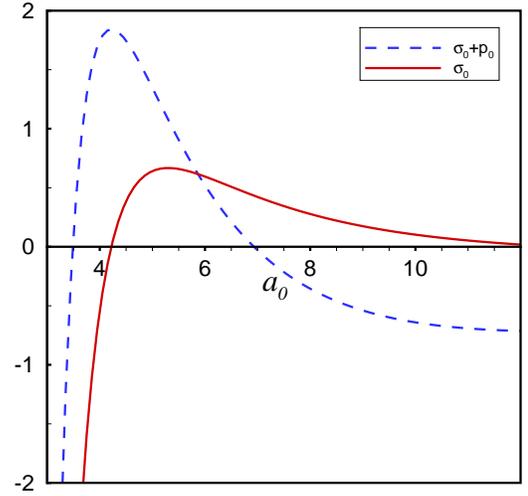}}
\caption{Specific case: $\protect\alpha _{2}<0$ and $\protect\alpha _{3}>0$. 
$\protect\sigma _{0}$ (continuous) and $\protect\sigma _{0}+p_{0}$ (dashed)
versus $a_{0}$ for $\protect\alpha _{3}=1$, $m=100$, $q=3000$ and $n=8$.
These figures depict that for $\protect\alpha _{2}<0$ and $\protect\alpha %
_{3}>0$, the WEC can be satisfied. The plots also show that for $\protect%
\alpha _{3}>0$, the amount of normal matter increases as $\protect\alpha %
_{2}<0$ decreases.}
\label{f78}
\end{figure*}

\begin{figure*}[tbp]
\centering
{\includegraphics[width=.47\textwidth]{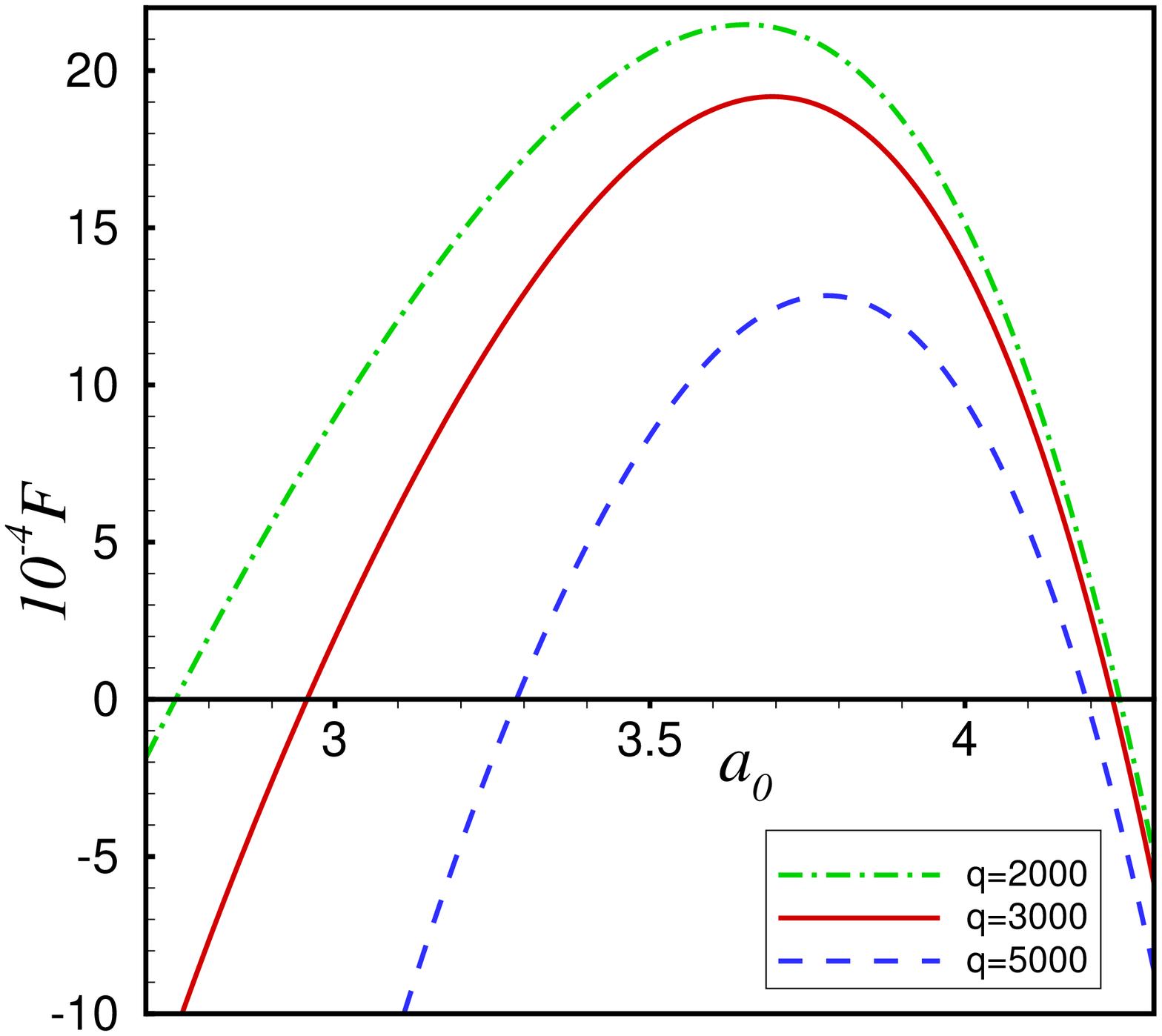}\qquad} {%
\includegraphics[width=.47\textwidth]{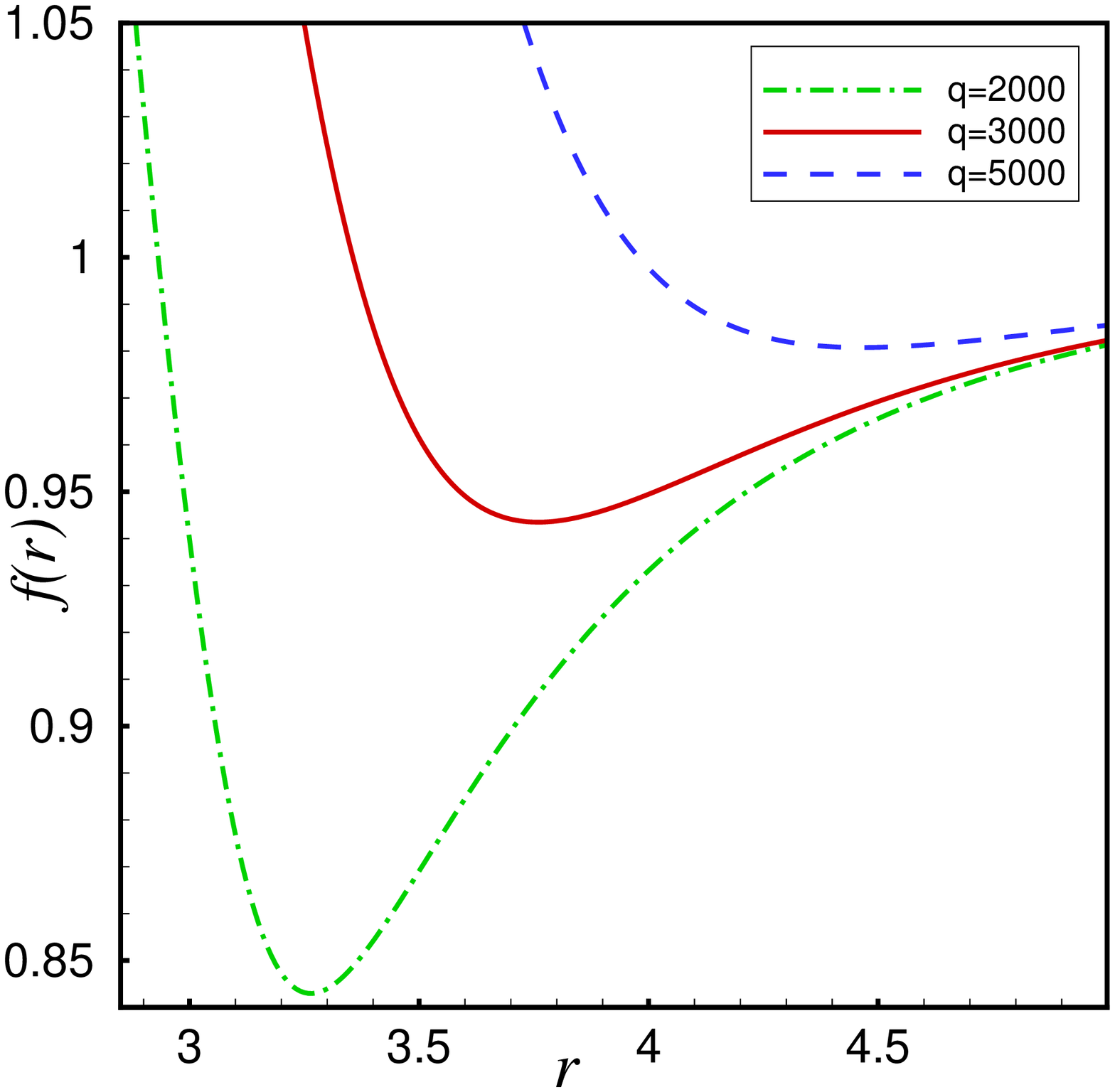}}
\caption{Right: $f(r)$ versus $r$ for $\protect\alpha _{2}=-0.6$, $\protect%
\alpha _{3}=0.2$, $n=8$, $m=300$ and $q=2000$, $3000$, $5000$ from down to
up respectively. Left: $10^{-4}\digamma $ versus $a_{0}$ for $f(r)$ plotted
in the right figure. These figures depict that for $\protect\alpha _{3}>0$
and $\protect\alpha _{2}<0$, the amount of normal matter decreases as $q$
increases.}
\label{f56}
\end{figure*}

For $\alpha _{3}>0$ and $\alpha _{2}<0$, $\digamma $ can be positive and
therefore the matter is normal. For this case, Figs. \ref{f34} and \ref{f78}
display the behaviour of $\sigma _{0}$ and $\sigma _{0}+p_{0}$ for $n=8$ and 
$9$ respectively. In these plots, there exist regions for $\sigma _{0}\geq 0$
(or $\digamma \geq 0$) and $\sigma _{0}+p_{0}\geq 0$ and therefore the WEC
and NEC are satisfied in this region. Since $0<f_{0}<1$,\ the factor $%
-15+10f_{0}-3f_{0}^{2}$ in Eq. (\ref{FF}) is negative and therefore the
amount of normal matter for negative $\alpha _{2}$ decreases as $\alpha
_{3}>0$ increases, as one can see in Figs. \ref{f34}-a) and \ref{f34}-b) for 
$\alpha _{3}=2$ and $\alpha _{3}=1$. In Figs. \ref{f78}-a) and \ref{f78}-b),
we show that the amount of normal matter increases as $\alpha _{2}<0$
decreases. Also, in this case the amount of normal matter decreases as the
charge $q$ increases (see Fig. \ref{f56}).

\subsection{Specific case: $\alpha_{2}>0$ and $\alpha_{3}<0$}

\begin{figure}[h]
\centering {\includegraphics[width=.47\textwidth]{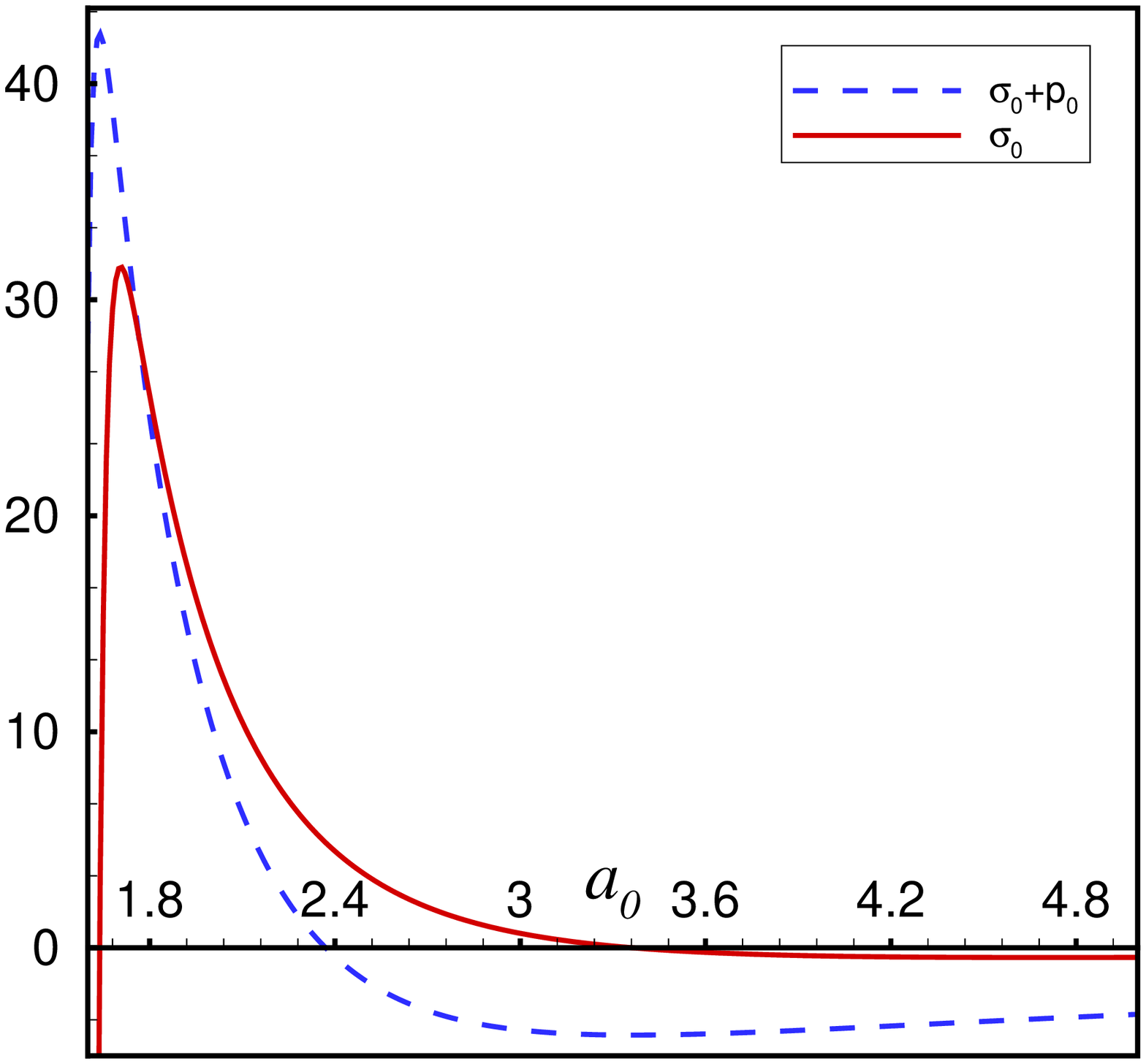}\qquad}
\caption{ Specific case: $\protect\alpha _{2}>0$ and $\protect\alpha _{3}<0$%
. $\protect\sigma _{0}$ (continuous) and $\protect\sigma _{0}+p_{0}$
(dashed) versus $a_{0}$ for $\protect\alpha _{2}=0.12$, $\protect\alpha %
_{3}=-0.1$, $m=10$, $q=1.4$ and $n=8$.}
\label{fim}
\end{figure}

Finally, we can see that for $\alpha _{2}>0$ and $\alpha _{3}<0$, $\digamma $
can be positive and therefore, for the first time, we have found thin-shell
wormholes threaded by normal matter on the shell in specific regions. In
this case, as depicted in Fig. \ref{fim}, there exist regions for $\sigma
_{0}\geq 0$ (or $\digamma \geq 0$) and $\sigma _{0}+p_{0}\geq 0$ and
therefore the WEC and NEC are satisfied. 

\section{STABILITY ANALYSIS}
\label{Stab}

In this section, we study the stability of the thin shell under small
perturbations preserving the symmetry of the wormhole configuration. The
dynamical evolution of the shell results from Eqs. (\ref{gesig0}) and (\ref%
{pre}) or by any of them and the conservation equation (\ref{eqco}). In
order to analyze the stability and to complete the system we use a cold
equation of state $p=p(\sigma )$ with $\eta =dp/d\sigma $. We consider small
radial perturbations around a static solution with radius $a_{0}$ and obtain
the equation of motion near the equilibrium solution. In this case, one may
write $p\simeq p_{0}+\eta _{0}(\sigma -\sigma _{0})$, where $p_{0}$, $\sigma
_{0}$ and $\eta _{0}$ are the transverse pressure, surface energy density
and $dp/d\sigma $ at the static solution $a=a_{0}$, respectively. Using this
linear equation of state, then Eq. (\ref{eqco}) yields the following
solution for the surface energy energy 
\begin{equation}
\sigma \left( a\right) =\left( \frac{\sigma _{0+}p_{0}}{1+\eta }\right)
\left( \frac{a_{0}}{a}\right) ^{(n-1)\left( 1+\eta \right) }+\frac{\eta
\sigma _{0-}p_{0}}{1+\eta } \,.  \label{EnD}
\end{equation}%
Substituting the energy density (\ref{EnD}) in Eq. (\ref{pre}), leads us to
the equation of motion, for the radius of the throat, given by 
\begin{eqnarray}
&&-\frac{60\pi a^{5}\sigma (a)}{(n-1)\sqrt{\dot{a}^{2}+f(a)}}=  \notag \\
&& 3\tilde{\alpha}_{3}\left\{8(1+\dot{a}^{2})^{2}+4[1-f(a)](1+\dot{a}%
^{2})+3[1-f(a)]^{2}\right\}  \notag \\
&&+15a^{4}+10\tilde{\alpha}_{2}a^{2}\left[ 2(1+\dot{a}^{2})+1-{f(a)}\right],
\label{eqgen}
\end{eqnarray}
where $\sigma (a)$ is given in Eq. (\ref{EnD}).

\begin{figure*}[tbp]
\centering
{\includegraphics[width=.40\textwidth]{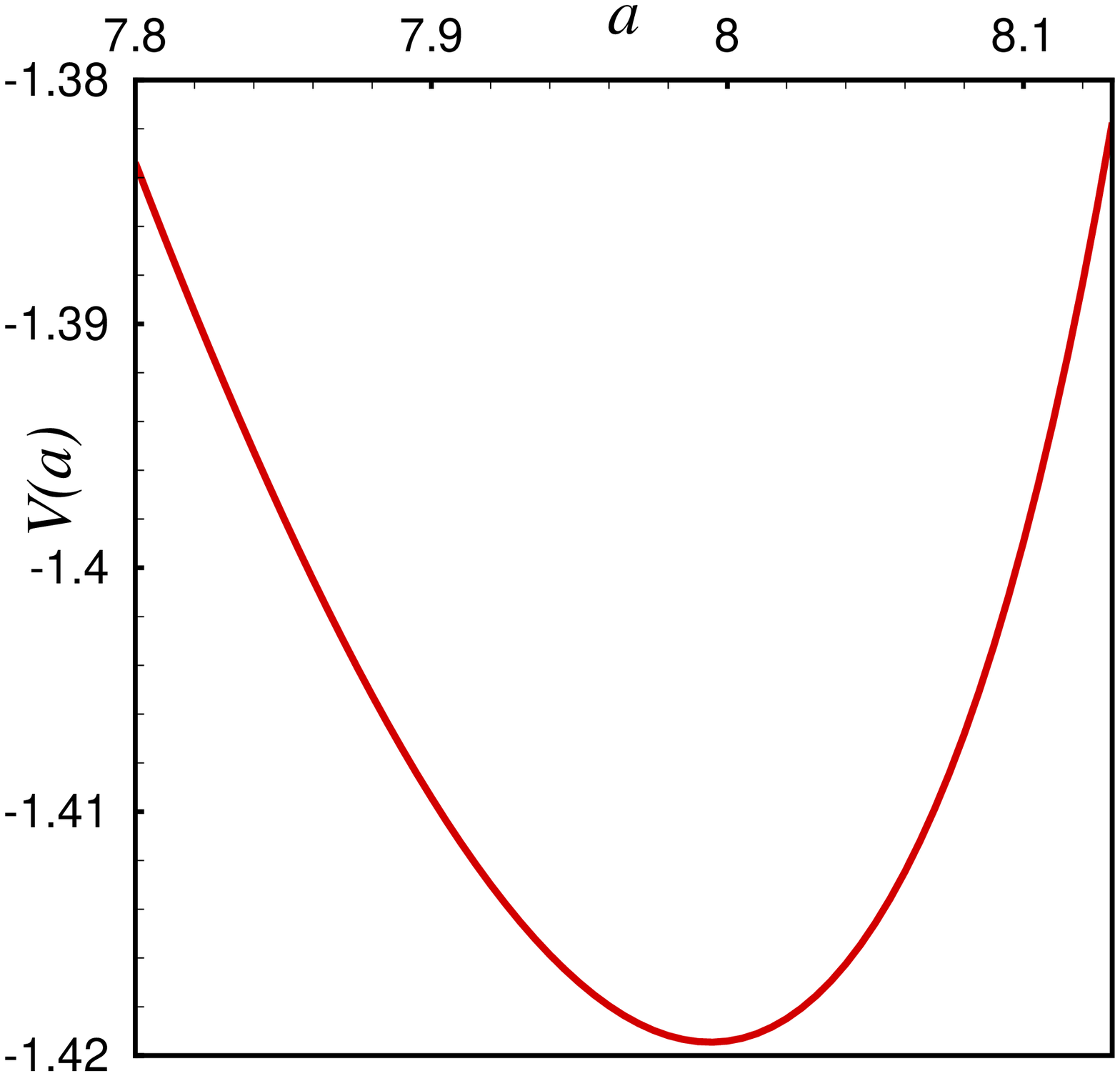}\qquad} {%
\includegraphics[width=.40\textwidth]{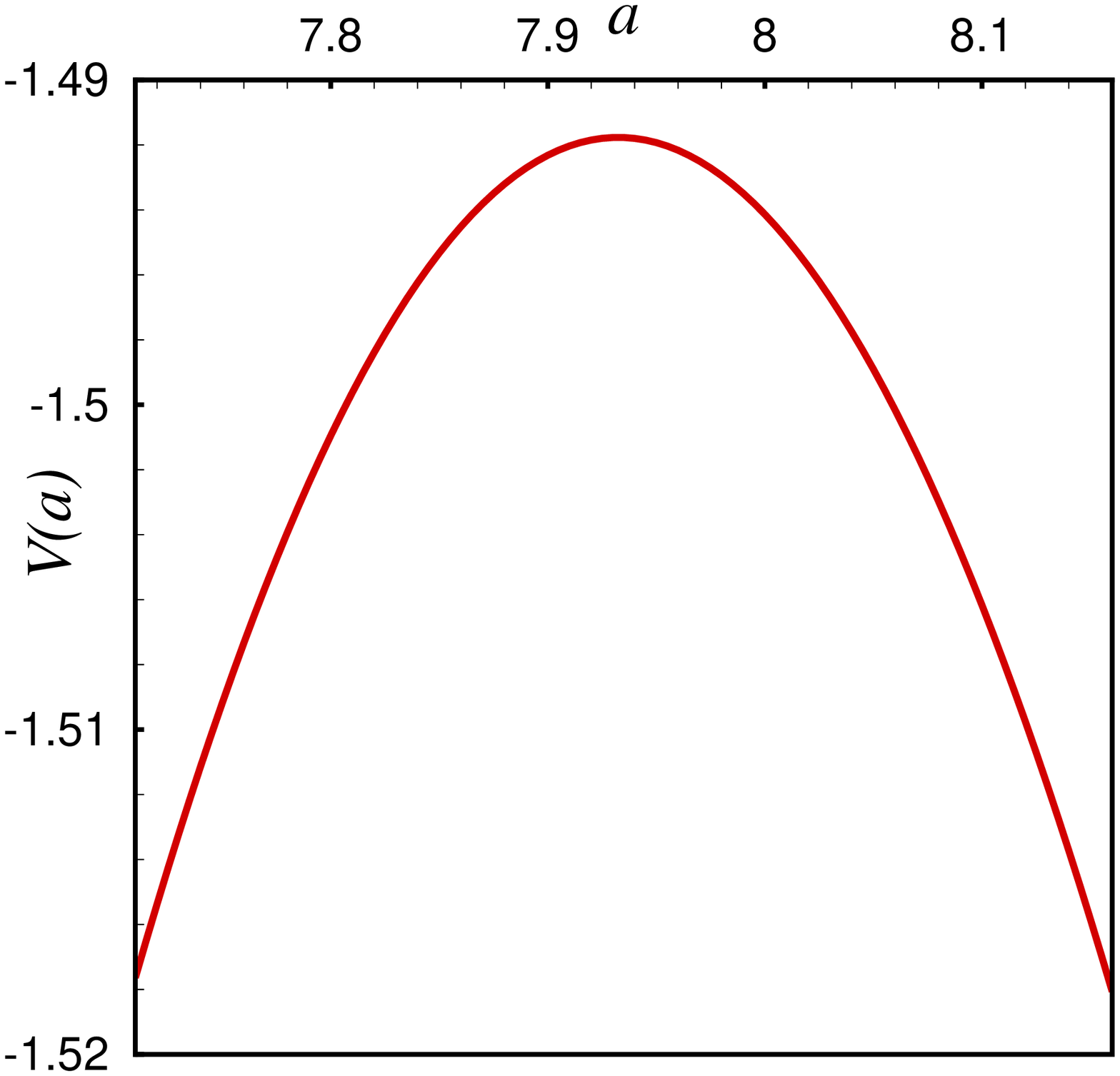}}
\caption{$V(a)$ versus $a$ with $\protect\alpha _{2}=-2$, $\protect\alpha %
_{3}=2$, $m=100$, $n=9$ and $q=3000$ for $\protect\eta =-3$ (left) and $%
\protect\eta =3$ (right). These figures depict that wormholes are stable for 
$\protect\eta =dp/d\protect\sigma <0$ and unstable for $\protect\eta >0$.}
\label{Stab1}
\end{figure*}

\begin{figure*}[tbp]
\centering
{\includegraphics[width=.40\textwidth]{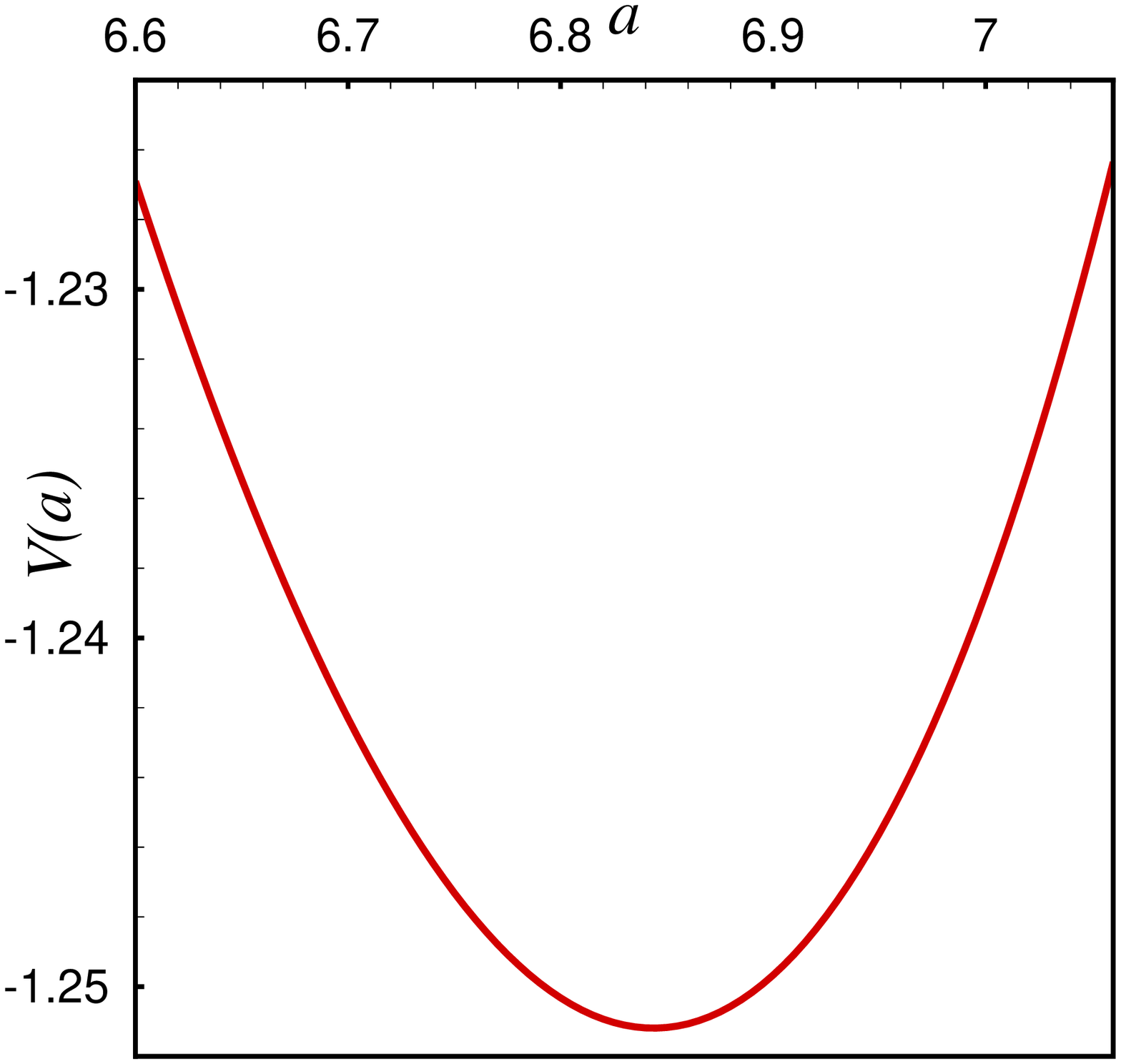}\qquad} {%
\includegraphics[width=.40\textwidth]{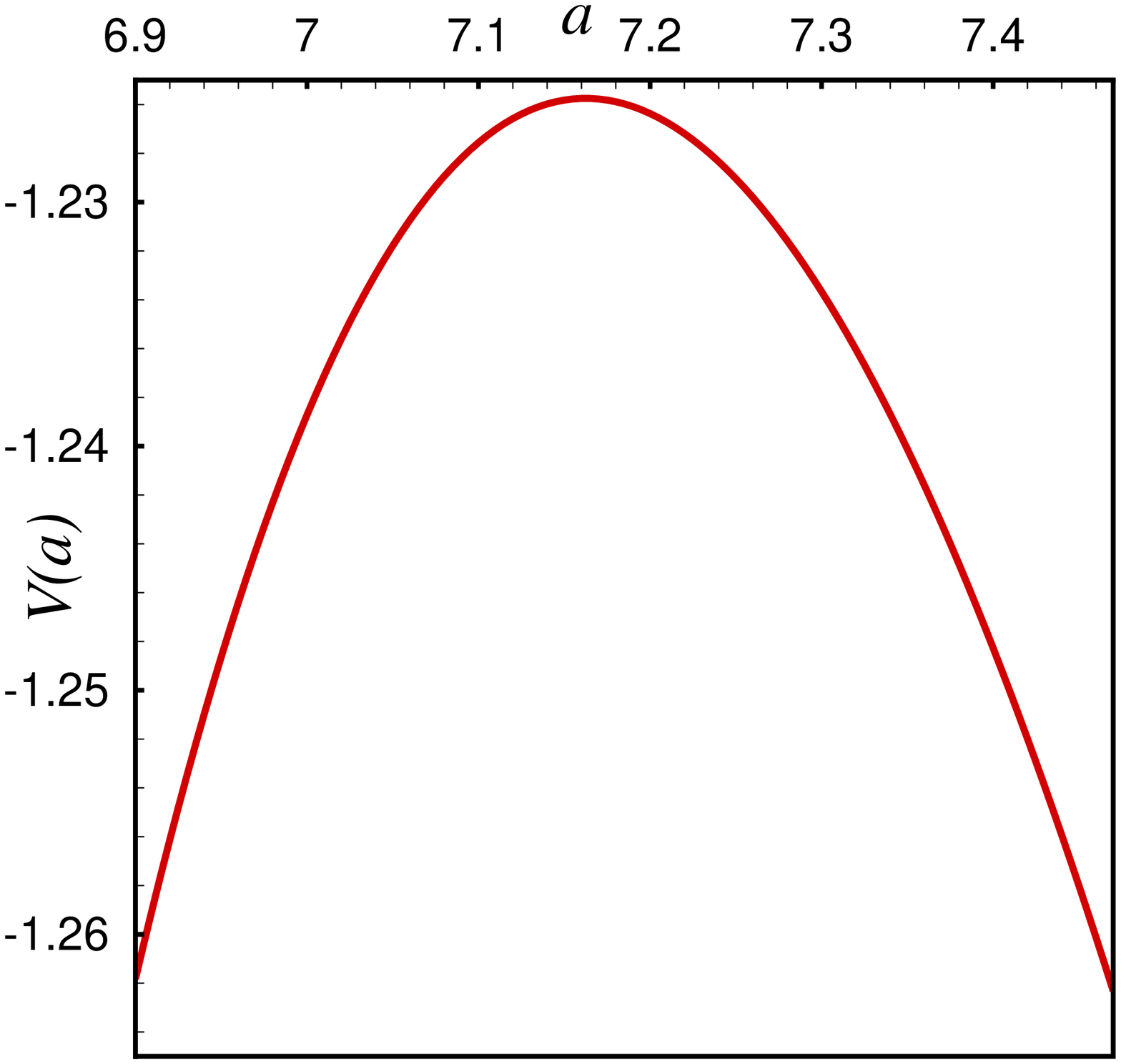}}
\caption{$V(a)$ versus $a$ with $\protect\alpha _{2}=-2$, $\protect\alpha %
_{3}=1$, $m=300$, $n=8$ and $q=3000$ for $\protect\eta =-2$ (left) and $%
\protect\eta =1$ (right). These figures depict that wormholes are stable for 
$\protect\eta =dp/d\protect\sigma <0$ and unstable for $\protect\eta >0$.}
\label{Stab3}
\end{figure*}

\begin{figure*}[tbp]
\centering
{\includegraphics[width=.40\textwidth]{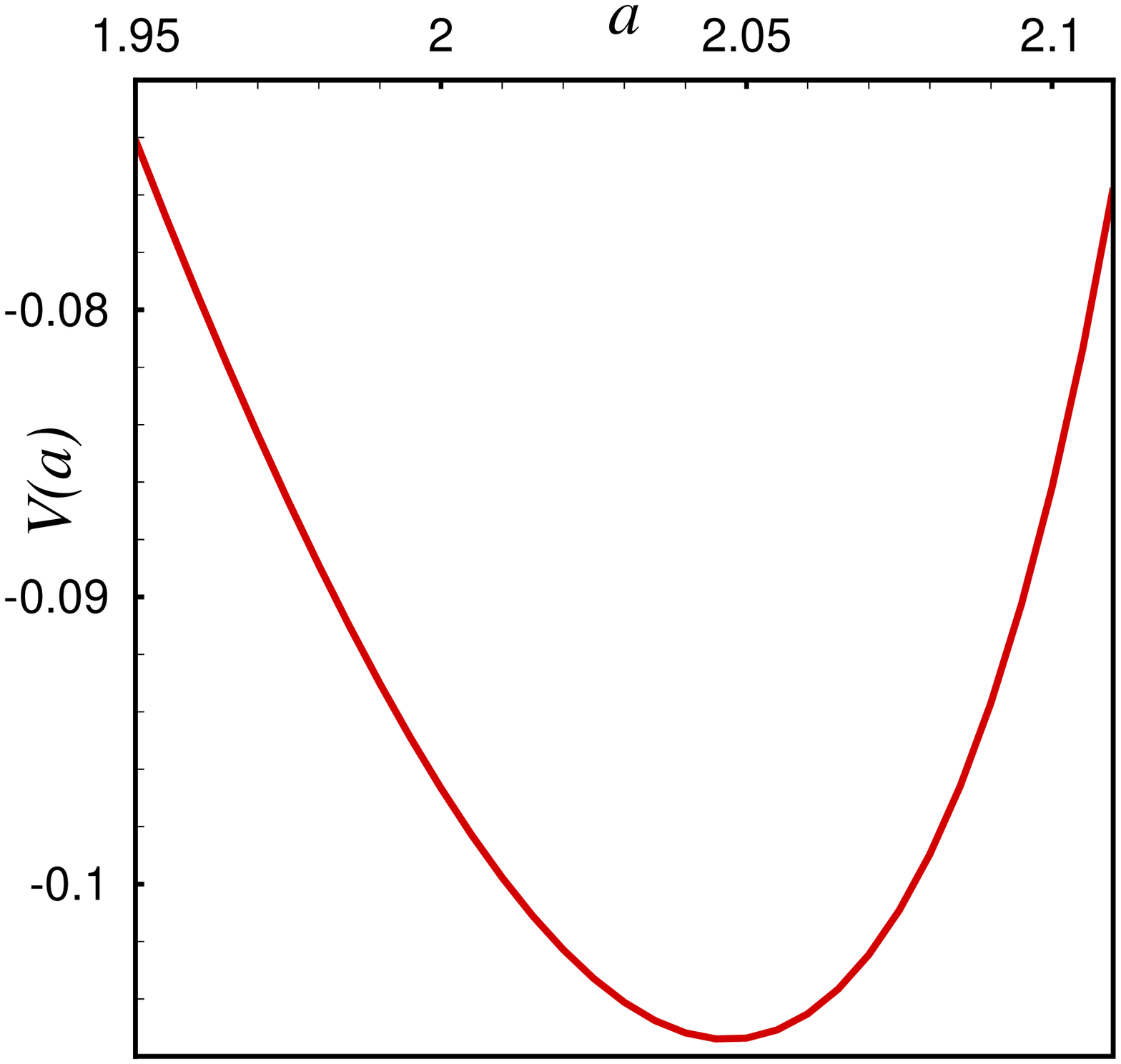}\qquad} {%
\includegraphics[width=.40\textwidth]{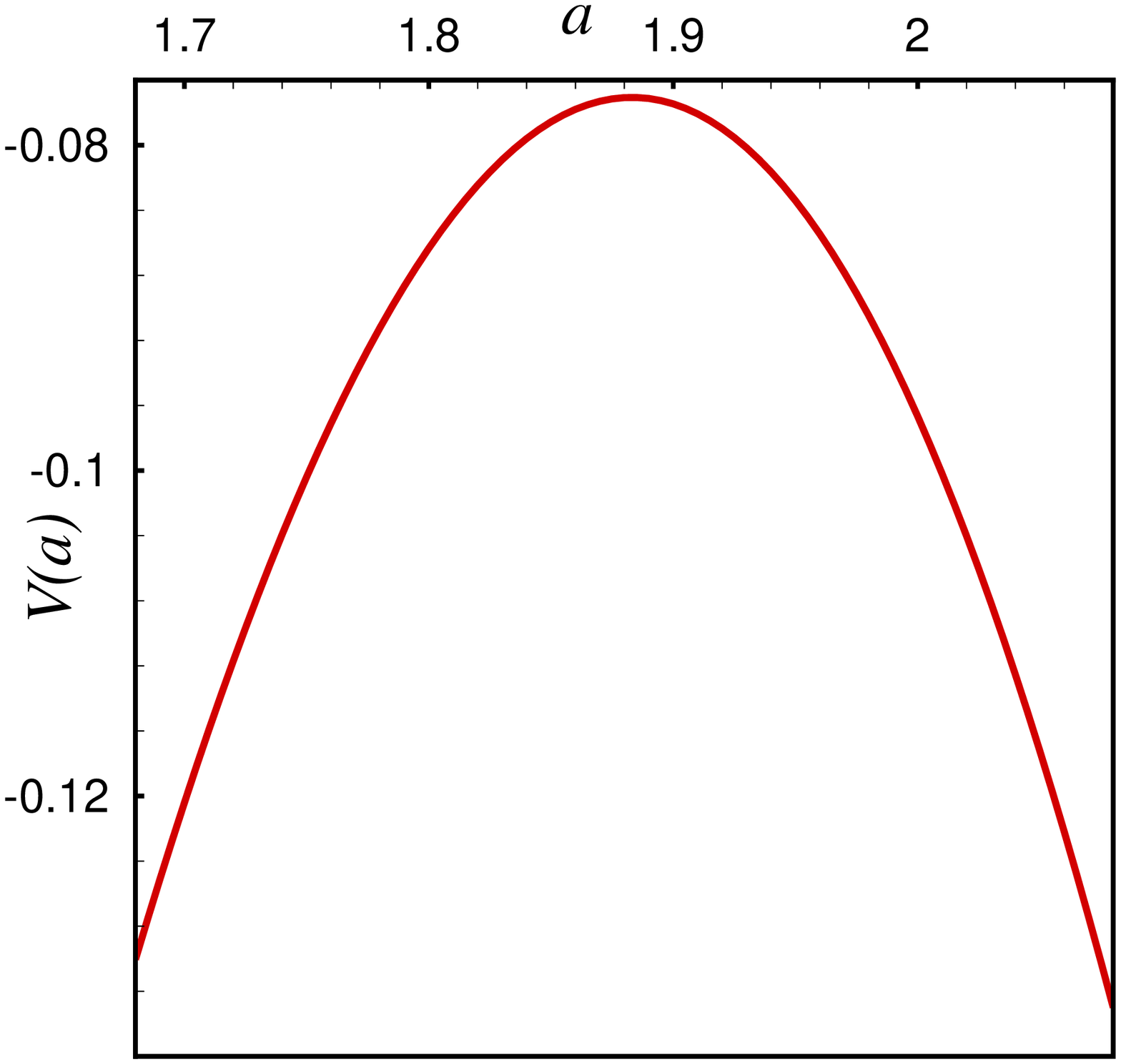}}
\caption{$V(a)$ versus $a$ with $\protect\alpha _{2}=0.12$, $\protect\alpha %
_{3}=-0.1$, $m=10$, $n=8$ and $q=1.4$ for $\protect\eta =-1.5$ (left) and $%
\protect\eta =1.5$ (right). These figures depict that wormholes are stable
for $\protect\eta =dp/d\protect\sigma <0$ and unstable for $\protect\eta >0$.}
\label{Stab2}
\end{figure*}

The dynamics of the wormhole throat is determined by a single equation with
the form of $\dot{a}^{2}=-V(a)$. However, in third order Lovelock gravity,
one encounters a fifth order algebraic equation for $\dot{a}^{2}$ and
unfortunately one cannot obtain an exact result, but, one may perform the
linearized stability analysis numerically. More specifically, one may solve
Eq. (\ref{eqgen}) for $\dot{a}^{2}$ numerically and obtain the potential $%
V(a)$ in the equation of motion. Then, the wormhole with radius $a_{0}$ is
linearly stable provided the potential $V(a)$ is minimum at $a=a_{0}$ where $%
a(\tau )$ will oscillate about $a_{0}$ and stabilize the wormhole. Numerical
calculations are shown in Figs. \ref{Stab1}-\ref{Stab2} for $n=8$ and $n=9$.
In the analysis, we have chosen the regions $a_{0}$ for which matter is
normal. As depicted by the figures, for the specific parameter ranges, the
wormholes are stable for negative values for $\eta $ and unstable for
positive $\eta =dp/d\sigma $. We emphasize that these results are in
agreement with those presented in \cite{meh}, where it was shown through
numerical calculations that wormholes are stable provided the parameter $%
\eta $ is negative and the throat radius $a_{0}$ is adequately chosen.

\section{SUMMARY AND DISCUSSION}\label{Concl}

Motivated by the fact that string theory proposes higher dimensional
gravitational objects, we have considered higher-dimensional thin-shell
wormholes in the context of third order Lovelock gravity. More specifically,
asymptotically flat charged thin-shell wormholes of third order Lovelock
gravity in higher dimensions were constructed, taking into account the
cut-and-paste technique. We extended the analysis outlined in \cite{meh}, by
providing the explicit form of the surface energy-momentum tensor in terms
of the extrinsic curvature and Riemann tensor for all higher dimensions,
explored the issue of the energy conditions and the amount of normal matter
that supports these thin-shell wormholes. Our analysis shows that for
negative second order and positive third-order Lovelock coefficients, there
are thin-shell wormhole solutions that respect the weak energy condition. In
this case, the amount of normal matter increases as the third-order Lovelock
coefficient decreases. We also found novel solutions which possess specific
regions where the energy conditions are satisfied for the case of a positive
second order and negative third-order Lovelock coefficients. Furthermore, we
presented the generalized junction conditions, and analysed the stability
analysis for the solutions found, in arbitrary dimensions. This enabled us
to study the effects of the third order Lovelock term in higher
dimensionality and the satisfaction of the WEC and stability of thin shell
wormhole solutions. In the latter context, a linear stability analysis in
higher dimensions around the static solutions was carried out, by
considering a specific cold equation of state, and a wide range of stability
regions were found.

\acknowledgements
MRM acknowledges Research Institute for Astronomy \&
Astrophysics of Maragha (RIAAM), Iran, for financial
support.
FSNL acknowledges financial support of the Funda\c{c}\~{a}o para a Ci\^{e}%
ncia e Tecnologia through an Investigador FCT Research contract, with
reference IF/00859/2012, and the grants EXPL/FIS-AST/1608/2013,
PEst-OE/FIS/UI2751/2014 and UID/FIS/04434/2013.

\appendix

\section{Action and spherically symmetric solutions}\label{app:I}

We reproduce the action of third order Lovelock gravity, in the presence of an
electromagnetic field, here
\begin{equation}
I=\int d^{n+1}x\sqrt{-g}\left( \mathcal{L}_{1}+\alpha _{2}\mathcal{L}%
_{2}+\alpha _{3}\mathcal{L}_{3}-F_{\mu \nu }F^{\mu \nu }\right) \,,
\label{Act1b}
\end{equation}%
where, as mentioned in Sec. \ref{Sol}, $\alpha _{2}$ and $\alpha _{3}$ are the second (Gauss-Bonnet) and
third order Lovelock coefficients. The term $\mathcal{L}_{1}=R$ is the
Einstein-Hilbert Lagrangian, $\mathcal{L}_{2}$ is the Gauss-Bonnet
Lagrangian given by 
\begin{equation}
\mathcal{L}_{2}=R_{\mu \nu \gamma \delta }R^{\mu \nu \gamma \delta }-4R_{\mu
\nu }R^{\mu \nu }+R^{2}\,,
\end{equation}%
and the third order Lovelock Lagrangian is defined as 
\begin{eqnarray}
\mathcal{L}_{3} &=&2R^{\mu \nu \sigma \kappa }R_{\sigma \kappa \rho \tau }R_{%
\phantom{\rho \tau }{\mu \nu }}^{\rho \tau }+8R_{\phantom{\mu \nu}{\sigma
\rho}}^{\mu \nu }R_{\phantom {\sigma \kappa} {\nu \tau}}^{\sigma \kappa }R_{%
\phantom{\rho \tau}{ \mu \kappa}}^{\rho \tau }  \notag \\
&&+24R^{\mu \nu \sigma \kappa }R_{\sigma \kappa \nu \rho }R_{%
\phantom{\rho}{\mu}}^{\rho }+3RR^{\mu \nu \sigma \kappa }R_{\sigma \kappa
\mu \nu }  \notag \\
&&+24R^{\mu \nu \sigma \kappa }R_{\sigma \mu }R_{\kappa \nu }+16R^{\mu \nu
}R_{\nu \sigma }R_{\phantom{\sigma}{\mu}}^{\sigma }  \notag \\
&&-12RR^{\mu \nu }R_{\mu \nu }+R^{3}.
\end{eqnarray}%
In Lovelock gravity only terms with order less than $[(n+1)/2]$ (where $[x]$ is the integer part of $x$)
contribute to the field equations, the rest being total derivatives in the
action. Therefore one should consider $n\geq 6$ in the case of
third order Lovelock gravity so that the effects of third order Lovelock
gravity appear.

The $(n+1)$-dimensional static solution of action (\ref{Act1b}) is given in
the form, which we rewrite for self-completeness,
\begin{equation}
ds^{2}=-f(r)dt^{2}+\frac{dr^{2}}{f(r)}+r^{2}d\Sigma _{n-1}^{2},  \label{metb}
\end{equation}%
where $d\Sigma _{n-1}^{2}$ is the metric of a $(n-1)$-dimensional unit
sphere and the metric function $f(r)$ is given by%
\begin{equation}
f(r)=1-r^{2}\psi (r).  \label{f(r)b}
\end{equation}

Substituting the electromagnetic potential (\ref{A}) and the metric (\ref{metb}) into the action (\ref{Act1b}) and varying the latter with respect to
the metric function $f(r)$, one obtains the equation of motion as \cite%
{amira} 
\begin{equation}
\tilde{\alpha}_{3}\psi ^{3}(r)+\tilde{\alpha}_{2}\psi ^{2}(r)+\psi (r)-\frac{%
m}{r^{n}}+\frac{q^{2}}{r^{2(n-1)}}=0,  \label{Cubic}
\end{equation}%
where, for notational simplicity, we define the following coefficients 
\begin{eqnarray}
\tilde{\alpha}_{2}&=&(n-3)(n-2)\alpha _{2}\,, \\
\tilde{\alpha}_{3}&=&(n-5)(n-4)(n-3)(n-2)\alpha _{3}\,.
\end{eqnarray}
The three solutions of Eq. (\ref{Cubic}) are%
\begin{align}
& \psi _{1}(r)=-\frac{\tilde{\alpha}_{2}}{3\tilde{\alpha}_{3}}+\delta
+u\,\delta ^{-1},  \label{Sol1} \\
& \psi _{2}(r)=-\frac{\tilde{\alpha}_{2}}{3\tilde{\alpha}_{3}}+-\frac{1}{2}%
(\delta +u\,\delta ^{-1})+i\frac{\sqrt{3}}{2}(\delta -u\,\delta ^{-1}),
\label{Sol2} \\
& \psi _{3}(r)=-\frac{\tilde{\alpha}_{2}}{3\tilde{\alpha}_{3}}-\frac{1}{2}%
(\delta +u\,\delta ^{-1})-i\frac{\sqrt{3}}{2}(\delta -u\,\delta ^{-1}),
\label{Sol3}
\end{align}%
where 
\begin{gather*}
\delta =\left( v+\sqrt{v^{2}-u^{3}}\right) ^{1/3},\text{ \ \ \ \ }u=\frac{%
\tilde{\alpha}_{2}^{2}-3\tilde{\alpha}_{3}}{9\tilde{\alpha}_{3}^{2}} \\
v=\frac{9\tilde{\alpha}_{2}\tilde{\alpha}_{3}-2\tilde{\alpha}_{2}^{3}}{54%
\tilde{\alpha}_{3}^{3}}-\frac{1}{2\tilde{\alpha}_{3}}\left[ \frac{q^{2}}{%
r^{2(n-1)}}-\frac{m}{r^{n}}\right] .
\end{gather*}%
All of the three roots of Eqs. (\ref{Sol1})-(\ref{Sol3}) allow real values
in appropriate ranges of $\tilde{\alpha}_{2}$ and $\tilde{\alpha}_{3}$. For
instance, $\psi_{1}(r)$ is real provided $u^{3}<v^{2}$ and $\psi _{2}(r)$
and $\psi _{3}(r)$ are real for $u^{3}>v^{2}$.

In this work, we will only consider $\psi _{1}$ and study some properties of
the metric function $f(r)$. Note that $f(r)=1-r^{2}\psi _{1}$ presents an
asymptotically flat black hole with two horizons provided $q<$ $q_{\mathrm{%
ext}}$, an extreme black hole with one horizon if $q=$ $q_{\mathrm{ext}}$
and a naked singularity if $q>q_{\mathrm{ext}}$, where $q_{\mathrm{ext}}$ is
related to $r_{\mathrm{ext}}$ through the following relationship 
\begin{equation}
\lbrack (n-6)\tilde{\alpha}_{3}+(n-4)\tilde{\alpha}_{2}r_{\mathrm{ext}%
}^{2}+(n-2)r_{\mathrm{ext}}^{4}]-\frac{(n-2)q_{\mathrm{ext}}^{2}}{r_{\mathrm{%
ext}}^{2(n-4)}}=0.
\end{equation}%
For the special case of $\tilde{\alpha}_{2}^{2}=3\tilde{\alpha}_{3}=\alpha
^{2}$, the function $\psi _{1}$ reduces to 
\begin{equation}
\psi _{1}(r)=-\frac{1}{\alpha }\left\{1-\left[ 1+3\alpha \left( \frac{m}{%
r^{n}}-\frac{q^{2}}{r^{2(n-1)}}\right) \right] ^{1/3}\right\}.  \label{F7sp}
\end{equation}%
In studying wormholes, we consider the region outside the horizon $r>r_{h}$,
where $r_{h}$ is the largest real root of $f(r)=0$ or $\psi (r)=1/r^{2}$.

\section{Junction conditions in Lovelock gravity}
\label{Thin}

In order to study these geometries, we first briefly introduce the junction
conditions in Lovelock gravity. Let us consider two manifolds $\mathcal{M}%
_{+} $ and $\mathcal{M}_{-}$ of $\mathcal{M}$ separated by a hypersurface $%
\Sigma $, and denote its two sides by $\Sigma _{\pm }$. In this case, we
must add boundary terms to the action so that the variation of the action
with respect to the metric is well-defined. The appropriate boundary terms
are \cite{rubm} 
\begin{align}
I_{\mathrm{sur}}=& \frac{1}{8\pi }\int_{\partial \mathcal{M}}d^{n}x\sqrt{%
-\gamma }\left[ K+2\alpha _{2}\left( J-2\hat{G}_{ab}^{(1)}K^{ab}\right)
\right.  \notag \\
&+ 3\alpha _{3}\left( P-2\hat{G}_{ab}^{(2)}K^{ab}-12\hat{R}_{ab}J^{ab}+2\hat{%
R}J\right.  \notag \\
& \left. \left. -4\hat{R}_{abcd}\left(
2K^{ac}K_{e}^{b}K^{ed}-KK^{ac}K^{bd}\right) \right) \right] ,
\end{align}%
where $J$ and $P$ are the traces of 
\begin{equation}
J_{ab}=\frac{1}{3}%
(2KK_{ac}K_{b}^{c}+K_{cd}K^{cd}K_{ab}-2K_{ac}K^{cd}K_{db}-K^{2}K_{ab}),
\end{equation}%
and 
\begin{eqnarray}
P_{ab} &=&\frac{1}{5}\{[K^{4}-6K^{2}K^{cd}K_{cd}+8KK_{cd}K_{e}^{d}K^{ec} 
\notag \\
&&-6K_{cd}K^{de}K_{ef}K^{fc}+3(K_{cd}K^{cd})^{2}]K_{ab}  \notag \\
&&-(4K^{3}-12KK_{ed}K^{ed}+8K_{de}K_{f}^{e}K^{fd})K_{ac}K_{b}^{c}  \notag \\
&&-24KK_{ac}K^{cd}K_{de}K_{b}^{e}+24K_{ac}K^{cd}K_{de}K^{ef}K_{bf}  \notag \\
&&+\left( 12K^{2}-12K_{ef}K^{ef}\right) K_{ac}K^{cd}K_{db}\}\,,
\end{eqnarray}
respectively, where $K_{ab}$ is the extrinsic curvature.

The tensors $\hat{G}_{ab}^{(1)}$ and $\hat{G}_{ab}^{(2)}$ are respectively
the Einstein and Gauss-Bonnet tensors associated with the induced metric $%
\gamma _{ab}$. Then, the variation of the total action $I+I_{\mathrm{sur}}$
with respect to $\gamma _{ab}$ provides the surface energy-momentum tensor $%
\mathcal{T}_{a}^{b}$. The discontinuous change of the surface
energy-momentum tensor on the boundary $\Sigma $ or junction condition can
be written as%
\begin{equation}
\mathcal{S}_{ab}=(\mathcal{T}_{+})_{ab}-(\mathcal{T}_{-})_{ab},
\label{explicit junction ij}
\end{equation}%
where the symmetric tensor $\mathcal{T}_{ab}$ is given by 
\begin{widetext}
\begin{eqnarray}
\mathcal{T}_{ab}&=&\frac{1}{8\pi }\{K_{ab}-K\gamma _{ab}+2\alpha
_{2}[3J_{ab}-J\gamma _{ab}-2\hat{G}_{(a}{}^{c}K_{b)c}+2\hat{R}_{ab}K-K_{ab}%
\hat{R}+2K_{cd}\hat{G}^{cd}\gamma _{ab}-2K^{cd}\hat{R}_{acbd}]  \notag \\
&&+3\alpha _{3}\{5P_{ab}-P\gamma _{ab}+2K\hat{G}_{ab}^{(2)}+\mathcal{L}%
_{2}(K_{ab}+K\gamma _{ab})+4J\hat{R}_{ab}-24J_{(a}{}^{c}\hat{R}_{b)c}+8K^{cd}%
\hat{R}_{ac}\hat{R}_{bd}-8KK_{a}{}^{c}K_{b}{}^{d}\hat{R}_{cd}  \notag \\
&&+8KK_{ab}K^{cd}\hat{R}_{cd}-8K^{cd}\hat{R}_{ab}\hat{R}_{cd}+16K_{(a}{}^{c}%
\hat{R}_{b)}{}^{d}\hat{R}_{cd}+16K_{(a}{}^{c}K_{b)}{}^{d}K_{d}{}^{e}\hat{R}%
_{ce}-8K_{ab}K_{c}{}^{e}K^{cd}\hat{R}_{de}+6J_{ab}\hat{R}  \notag \\
&&-8K_{(a}{}^{c}\hat{R}_{b)c}\hat{R}-12J^{cd}\hat{R}_{acbd}-4K^{cd}\hat{R}%
\hat{R}_{acbd}-8K_{a}{}^{c}K_{bc}K^{de}\hat{R}%
_{de}+16K_{(a}{}^{c}K_{c}{}^{d}K^{ef}\hat{R}%
_{b)edf}+16K_{(a}{}^{c}K_{d}{}^{f}K^{de}\hat{R}_{b)ecf}  \notag \\
&&+16K_{(a}{}^{c}\hat{R}^{de}\hat{R}_{b)dce}-16KK_{(a}{}^{c}K^{de}\hat{R}%
_{b)dce}-16K^{cd}\hat{R}_{(a}{}^{e}\hat{R}_{b)cde}-4K^{cd}\hat{R}_{ac}{}^{ef}%
\hat{R}_{bdef}+8K^{cd}\hat{R}_{c}{}^{e}\hat{R}_{aebd}  
  \notag \\
&&+8K^{cd}\hat{R}%
_{c}{}^{e}\hat{R}_{adbe}-8K^{cd}\hat{R}_{a}{}^{e}{}_{c}{}^{f}\hat{R}_{bedf}-8K_{(a}{}^{c}\hat{R}%
_{b)}{}^{def}\hat{R}_{cdef}+8K_{a}{}^{c}K_{b}{}^{d}K^{ef}\hat{R}%
_{cedf}-4K_{ab}K^{cd}K^{ef}\hat{R}_{cedf}
    \notag \\
&& +8K^{cd}\hat{R}_{a}{}^{e}{}_{b}{}^{f}\hat{R}_{cedf}   +2\gamma _{ab}(\hat{G}_{cd}^{(2)}K^{cd}+6J^{cd}\hat{R}_{cd}-J\hat{R}%
+2KK^{cd}K^{ef}\hat{R}_{cedf}-4K_{c}{}^{e}K^{cd}K^{fh}\hat{R}_{dfeh})\}.
\end{eqnarray}%
\end{widetext}


\begin{thebibliography}{99}
\bibitem{MT} M. S. Morris and K. S. Thorne, Am. J. Phys. \textbf{56}, 395
(1986).

\bibitem{MTY} M. S. Morris, K. S. Thorne, and U. Yurtsever, Phys. Rev. Lett. 
\textbf{61}, 1446 (1988); M. Visser, Lorentzian Wormholes: From Einstein to
Hawking (American Institute of Physics, New York, 1995).

\bibitem{viskardad} M. Visser, S. Kar and N. Dadhich, Phys. Rev. Lett. 
\textbf{90}, 201102 (2003).

\bibitem{MartRich} M. G. Richarte, \ Phys. Rev. D \textbf{82}, 044021 (2010).

\bibitem{Deh1} M. H. Dehghani and Z. Dayyani, Phys. Rev. D \textbf{79},
064010 (2009).

\bibitem{wecsat} M. Kord Zangeneh, F. S. N. Lobo and N. Riazi, Phys. Rev. D 
\textbf{90}, 024072 (2014); M. R. Mehdizadeh, M. Kord Zangeneh and F. S. N.
Lobo, Phys. Rev. D \textbf{91}, 084004 (2015).


\bibitem{Harko:2013yb}  T.~Harko, F.~S.~N.~Lobo, M.~K.~Mak and
S.~V.~Sushkov,  
Phys.\ Rev.\ D \textbf{87}, no. 6, 067504 (2013).  

\bibitem{Vis} M. Visser, Phys. Rev. D \textbf{39,} 3182 (1989); Nucl. Phys.
B \textbf{328}, 203 (1989).

\bibitem{VisP} E. Poisson and M. Visser, Phys. Rev. D 52, 7318 (1995).

\bibitem{thin} S. W. Kim, Phys. Lett. A \textbf{166}, 13 (1992); F. S. N.
Lobo, Classical Quantum Gravity \textbf{21}, 4811 (2004); J. P. S. Lemos and
F. S. N. Lobo, Phys. Rev. D \textbf{69}, 104007 (2004); E. F. Eiroa and C.
Simeone, ibid. \textbf{70}, 044008 (2004); E. F. Eiroa and C. Simeone, ibid. 
\textbf{71}, 127501 (2005); F. Rahaman, M. Kalam and S. Chakraborty, Gen.
Relativ. Gravit. \textbf{38}, 1687 (2006); C. Bejarano, E. F. Eiroa and C.
Simeone, Phys. Rev. D \textbf{75}, 027501 (2007); F. Rahaman, M. Kalam, and
S. Chakraborty, Int. J. Mod. Phys. D \textbf{16}, 1669 (2007); F. Rahaman,
M. Kalam, K. A. Rahman and S. Chakraborty, Gen. Relativ. Gravit. \textbf{39}%
, 945 (2007); E. Gravanis and S. Willison, Phys. Rev. D \textbf{75}, 084025
(2007); M. G. Richarte and C. Simeone, Int. J. Mod. Phys. D \textbf{17},
1179 (2008); S. H. Mazharimousavi, M. Halilsoy, Z. Amirabi, Eur. Phys. J. C 
\textbf{74}, 2889 (2014).

\bibitem{lemo11} M. Thibeault, C. Simeone, and E. F. Eiroa, Gen. Rel. Grav. 
\textbf{38}, 1593 (2006); F. Rahaman, M. Kalam, and S. Chakraborty, Gen.
Rel. Grav. \textbf{38}, 1687 (2006); F. Rahaman, M. Kalam and S.
Chakraborty, Int. J. Mod. Phys. D \textbf{16}, 1669 (2007); E. Gravanis and
S. Willison, Phys. Rev. D \textbf{75}, 084025 (2007).

\bibitem{bej} E. F. Eiroa and C. Simeone, Phys. Rev. D \textbf{71}, 127501
(2005); C. Bejarano and E. F. Eiroa, Phys. Rev. D \textbf{84}, 064043
(2011); S. H. Mazharimousavi, M. Halilsoy and Z. Amirabi, Phys. Lett. A 
\textbf{375, }231 (2011).

\bibitem{yueca} X. Yue and S. Gao, Phys. Lett. A \textbf{375}, 2193 (2011);
Francisco S. N. Lobo and Miguel A. Oliveira, Phys. Rev. D \textbf{81},
067501 (2010).

\bibitem{cyl} M. G. Richarte Phys. Rev. D \textbf{87},067503 (2013).

\bibitem{brad} P. R. Brady, J. Louko and E. Poisson, Phys. Rev. D \textbf{44}%
, 1891 (1991); M. Ishak and K. Lake, Phys. Rev. D \textbf{65}, 044011
(2002); S. M. C. V. Gon%
\c{}%
calves, Phys. Rev. D \textbf{66}, 084021 (2002); F. S. N. Lobo and P.
Crawford, Class. Quantum Grav. \textbf{22}, 4869 (2005); E. F. Eiroa and C.
Simeone, Phys. Rev. D \textbf{83,} 104009 (2011); Mustapha Ishak and Kayll
Lake, Phys. Rev. D \textbf{65,} 044011 (2002); Ernesto F. Eiroa, Phys. Rev.
D \textbf{78}, 024018 (2008); N. M. Garcia, F. S. N. Lobo and M. Visser,
Phys. Rev. D \textbf{86, }044026 (2012); C. Bejarano and E. F. Eiroa, Phys.
Rev. D \textbf{84, }064043 (2011); M. Sharif and M. Azam, JCAP \textbf{1305, 
}025 (2013); Z. Amirabi, M. Halilsoy and S. H. Mazharimousavi, Phys. Rev. D 
\textbf{88}, 124023 (2013).

\bibitem{rah1} F. Rahaman, M. Kalam and S. Chakraborty, Gen. Relativ.
Gravit. \textbf{38}, 1687 (2006); J. P. S. Lemos and F. S. N. Lobo, Phys.
Rev. D \textbf{78}, 044030 (2008); G. A. S. Dias and J. P. S. Lemos, Phys.
Rev. D \textbf{82}, 084023 (2010); S. H. Mazharimousavi, M. Halilsoy, Z.
Amirabi, Class. Quant. Grav. \textbf{28}, 025004 (2011).

\bibitem{eir} E. F. Eiroa and C. Simeone, Phys. Rev. D \textbf{70}, 044008
(2004); E. F. Eiroa, Phys. Rev. D \textbf{78}, 024018 (2008); A. A. Usmani,
F. Rahaman, S. Ray, Sk. A. Rakib and Z. Hasan, Gen. Relativ. Gravit. \textbf{%
42}, 2901 (2010); P. K. F. Kuhfittig, Acta. Phys. Pol. B \textbf{41}, 2017
(2010); E. F. Eiroa and C. Simeone, Phys. Rev. D \textbf{81}, 084022 (2010);
M. G. Richarte, Phys. Rev. D \textbf{88}, 027507(2013); S. H.
Mazharimousavi, M. Halilsoy and Z. Amirabi, Phys. Rev. D \textbf{89}, 084003
(2014); K. A. Bronnikov, V. G. Krechet and J. P. S. Lemos, Phys. Rev. D 
\textbf{87}, 084060 (2013).

\bibitem{lovee} D. Lovelock, J. Math. Phys. (N.Y.) \textbf{12}, 498 (1971);
D. Lovelock, Aequationes mathematicae \textbf{4}, 127 (1970).

\bibitem{mazv} S. H. Mazharimousavi, M. Halilsoy and Z. Amirabi, Phys. Rev.
D \textbf{81}, 104002 (2010); Classical Quantum Gravity \textbf{28}, 025004
(2011).

\bibitem{meh} M. H. Dehghani and M. R. Mehdizadeh, Phys. Rev. D \textbf{85},
024024 (2012).

\bibitem{amira} Z. Amirabi, Phys. Rev. D \textbf{88}, 087503 (2013).

\bibitem{rubm} M. H. Dehghani and R. B. Mann, Phys. Rev. D \textbf{73},
104003 (2006); M. H. Dehghani, N. Bostani and A. Sheykhi, ibid. \textbf{73},
104013 (2006).
\end{thebibliography}
\end{document}